\documentclass[iop]{emulateapj}

\usepackage{apjfonts}
\usepackage{etoolbox}
\makeatletter
\patchcmd{\NAT@citex}
  {\@citea\NAT@hyper@{%
     \NAT@nmfmt{\NAT@nm}%
     \hyper@natlinkbreak{\NAT@aysep\NAT@spacechar}{\@citeb\@extra@b@citeb}%
     \NAT@date}}
  {\@citea\NAT@nmfmt{\NAT@nm}%
   \NAT@aysep\NAT@spacechar\NAT@hyper@{\NAT@date}}{}{}

\patchcmd{\NAT@citex}
  {\@citea\NAT@hyper@{%
     \NAT@nmfmt{\NAT@nm}%
     \hyper@natlinkbreak{\NAT@spacechar\NAT@@open\if*#1*\else#1\NAT@spacechar\fi}%
       {\@citeb\@extra@b@citeb}%
     \NAT@date}}
  {\@citea\NAT@nmfmt{\NAT@nm}%
   \NAT@spacechar\NAT@@open\if*#1*\else#1\NAT@spacechar\fi\NAT@hyper@{\NAT@date}}
  {}{}
\makeatother

\usepackage{ifthen}
\usepackage{natbib}
\usepackage{astjnl-notinemulateapj}
\bibpunct[, ]{(}{)}{,}{a}{}{,}
\usepackage{amssymb, amsmath}
\usepackage{appendix}

\usepackage[
bookmarks=true,                    
bookmarksnumbered=true,            
colorlinks=true,            
citecolor=blue,         
linkcolor=blue,     
menucolor=blue,         
urlcolor=blue,       
linkbordercolor={0 0 1},           
pdfborder={0 0 1},
frenchlinks=true]{hyperref}

\shorttitle{Secondary Eclipses of HAT-P-13\MakeLowercase{b}}
\shortauthors{Hardy {\em et al.} 2016}

\newcommand\degree{\degr}
\newcommand\degrees\degree

\DeclareSymbolFont{UPM}{U}{eur}{m}{n}
\DeclareMathSymbol{\umu}{0}{UPM}{"16}
\let\oldumu=\umu
\renewcommand\umu{\ifmmode\oldumu\else\math{\oldumu}\fi}
\newcommand\micro{\umu}
\renewcommand{\micron}{\micro m}
\newcommand\microns {\micron}

\let\oldsim=\sim
\renewcommand\sim{\ifmmode\oldsim\else\math{\oldsim}\fi}
\let\oldpm=\pm
\renewcommand\pm{\ifmmode\oldpm\else\math{\oldpm}\fi}
\newcommand\by{\ifmmode\times\else\math{\times}\fi}

\newcommand\tablebox[1]{\begin{tabular}[t]{@{}l@{}}#1\end{tabular}}
\newbox{\wdbox}
\renewcommand\c{\setbox\wdbox=\hbox{,}\hspace{\wd\wdbox}}
\renewcommand\i{\setbox\wdbox=\hbox{i}\hspace{\wd\wdbox}}

\newcount\timect
\newcount\hourct
\newcount\minct
\newcommand\now{\timect=\time \divide\timect by 60
         \hourct=\timect \multiply\hourct by 60
         \minct=\time \advance\minct by -\hourct
         \number\timect:\ifnum \minct < 10 0\fi\number\minct}

\newcommand\mctc{\multicolumn{2}{c}}

\catcode`@=11

\newcommand\comment[1]{}

\renewcommand\math[1]{$#1$}
\newcommand\mathshifton{\catcode`\$=3}
\newcommand\mathshiftoff{\catcode`\$=12}

\comment{the backslash is necessary}

\comment{alignment tab}

\let\atab=&
\newcommand\atabon{\catcode`\&=4}
\newcommand\ataboff{\catcode`\&=12}

\let\oldmsp=\sp
\let\oldmsb=\sb
\def\sp#1{\ifmmode
           \oldmsp{#1}%
         \else\strut\raise.85ex\hbox{\scriptsize #1}\fi}
\def\sb#1{\ifmmode
           \oldmsb{#1}%
         \else\strut\raise-.54ex\hbox{\scriptsize #1}\fi}
\newbox\@sp
\newbox\@sb
\def\sbp#1#2{\ifmmode%
           \oldmsb{#1}\oldmsp{#2}%
         \else
           \setbox\@sb=\hbox{\sb{#1}}%
           \setbox\@sp=\hbox{\sp{#2}}%
           \rlap{\copy\@sb}\copy\@sp
           \ifdim \wd\@sb >\wd\@sp
             \hskip -\wd\@sp \hskip \wd\@sb
           \fi
        \fi}
\def\msp#1{\ifmmode
           \oldmsp{#1}
         \else \math{\oldmsp{#1}}\fi}
\def\msb#1{\ifmmode
           \oldmsb{#1}
         \else \math{\oldmsb{#1}}\fi}
\def\supon{\catcode`\^=7}
\def\supoff{\catcode`\^=12}
\def\subon{\catcode`\_=8}
\def\suboff{\catcode`\_=12}
\def\supsubon{\supon \subon}
\def\supsuboff{\supoff \suboff}

\comment{
\let\oldmsp=\sp
\let\oldmsb=\sb
\renewcommand\sp[1]{\ifmmode
     \oldmsp{#1}%
   \else\strut\raise.85ex\hbox{\scriptsize #1}\fi}
\renewcommand\sb[1]{\ifmmode
     \oldmsb{#1}%
   \else\strut\raise-.54ex\hbox{\scriptsize #1}\fi}
\newcommand\msp[1]{\ifmmode
     \oldmsp{#1}
   \else \math{\oldmsp{#1}}\fi}
\newcommand\msb[1]{\ifmmode
     \oldmsb{#1}
   \else \math{\oldmsb{#1}}\fi}
\newcommand\supon{\catcode`\^=7}
\newcommand\supoff{\catcode`\^=12}
\newcommand\subon{\catcode`\_=8}
\newcommand\suboff{\catcode`\_=12}
\newcommand\supsubon{\supon \subon}
\newcommand\supsuboff{\supoff \suboff}
}

\newcommand\actcharon{\catcode`\~=13}
\newcommand\actcharoff{\catcode`\~=12}

\newcommand\paramon{\catcode`\#=6}
\newcommand\paramoff{\catcode`\#=12}

\comment{And now to turn us totally on and off...}

\newcommand\reservedcharson{\mathshifton \atabon \supsubon \actcharon
  \paramon}

\newcommand\reservedcharsoff{\mathshiftoff \ataboff
  \supsuboff \actcharoff \paramoff}

\catcode`@=12
\reservedcharsoff

\reservedcharson

\comment{ Must have ONLY ONE of these... trust these macros, they work

}
\reservedcharsoff

\actcharon

\comment{
\received{}
\revised{}
\accepted{}
}

\slugcomment{\tablebox{
ArXiv Preprint {\today} \now\\
Submitted to {\em ApJ}\/ 2015 November 5 23:30.\\
Accepted for publication by {\em ApJ} 2016 December 12.
\comment{
Accepted for publication by {\em ApJ} 2016 December 12. \\
Last revision: {\today} \now.}}}
\journalinfo{}

\reservedcharson
\begin{document}



\title {The Atmosphere and Interior Structure of {HAT-P-13\MakeLowercase{b}} from {\em Spitzer} Secondary Eclipses}

\author{Ryan A. Hardy\altaffilmark{1,2}}
\author{Joseph Harrington\altaffilmark{1,3}}
\author{Matthew R. Hardin\altaffilmark{1}}
\author{Nikku Madhusudhan\altaffilmark{4}}
\author{Thomas J. Loredo\altaffilmark{5}}
\author{Ryan C. Challener\altaffilmark{1}}
\author{Andrew S.D. Foster\altaffilmark{1}}
\author{Patricio E.\ Cubillos\altaffilmark{1,6}}
\author{Jasmina Blecic\altaffilmark{1}}

\email{email address: ryan.hardy@colorado.edu}

\affil{\sp{1}Planetary Sciences Group, Department of Physics, University of Central Florida\\
Orlando, FL 32816-2385}
\affil{\sp{2}Department of Aerospace Engineering Sciences, University of Colorado Boulder, Boulder, CO 80309, USA}
\affil{\sp{3}Max-Planck-Institut f\"{u}r Astronomie\\
K\"{o}nigstuhl 17, D-69117 Heidelberg, Germany}

\affil{\sp{4}Institute of Astronomy, University of Cambridge, Madingley Road, Cambridge CB3 0HA, UK.}
\affil{\sp{5}Department of Astronomy, Cornell University, Ithaca NY 14850, USA}
\affil{\sp{6}Space Research Institute, Austrian Academy of Sciences, Schmiedlstrasse 6, A-8042 Graz, Austria.}

\begin{abstract}

We present {\em Spitzer} secondary-eclipse observations of the hot Jupiter HAT-P-13 b in the 3.6 {\micron} and 4.5 {\micron} bands.  HAT-P-13 b inhabits a two-planet system with a configuration that enables constraints on the planet's second Love number, \math{k\sb{2}}, from precise eccentricity measurements, which in turn constrains models of the planet's interior structure. We exploit the direct measurements of \math{e \cos \omega} from our secondary-eclipse data and combine them with previously published radial velocity data to generate a refined model of the planet's orbit and thus an improved estimate on the possible interval for \math{k\sb{2}}.  We report eclipse phases of \math{0.49154 \pm 0.00080} and \math{0.49711 \pm 0.00083} and corresponding \math{e \cos \omega} estimates of \math{-0.0136 \pm 0.0013} and \math{-0.0048 \pm 0.0013}. Under the assumptions of previous work, our estimate of \math{k\sb{2}} of 0.81 {\pm} 0.10 is consistent with the lower extremes of possible core masses found by previous models, including models with no solid core. This anomalous result challenges both interior models and the dynamical assumptions that enable them, including the essential assumption of apsidal alignment. We also report eclipse depths of 0.081\% {\pm} 0.008\%  in the 3.6 {\micron} channel and 0.088 \% {\pm} 0.028 \% in the 4.5 {\micron} channel. These photometric results are non-uniquely consistent with solar-abundance composition without any thermal inversion.

\end{abstract}

\section*{}

\section{INTRODUCTION}

The G-type star HAT-P-13 hosts two planets, HAT-P-13 b and HAT-P-13 c, the discovery of which were reported by \citet{Bakos2009discovery}.  The smaller of the two, HAT-P-13 b, orbits close to its host star with a period of 2.91 days.  It has an estimated mass of 0.85 \math{M\sb{\rm J}} and an inflated radius of 1.28 \math{R\sb{\rm J}} \citep{Winn2010RV} and an estimated equilibrium temperature of 1626 \pm 42 K assuming zero Bond albedo and uniform heat redistribution.  Its orbit was found to have a non-negligible eccentricity of 0.013 \citep{Bakos2009discovery, Winn2010RV}.  The massive outer planet, HAT-P-13 c, occupies a highly eccentric orbit (\math{e = 0.66)} with a period of about 446 days.  The increased separation of planet c from its host makes observing a transit improbable and observation windows long and infrequent.  To date, a transit of planet c has not been observed.  

The HAT-P-13 system is unique in that its two-planet configuration allows for a measurement of the interior structure of HAT-P-13 b under equilibrium tide theory assuming a fixed-point eccentricity, i.e., assuming that all libration has been damped out and the rate at which the outer planet pumps the inner planet's eccentricity is balanced by the rate at which tidal deformation dissipates this energy. This requires the circularization timescale to be considerably shorter than the estimated age of the system of 5 Gyr. \cite{Mardling2010} calculates that this timescale is likely less than 1 Gyr for a reasonable range of damping coefficients.
Under this theory, a precise measurement of its eccentricity would directly yield the value of its second Love number, \math{k\sb{\rm{2b}}}, which describes how the planet's gravitational field changes in response to external potentials \citep{RagozzineWolf2009interiors}. Such a solution would provide constraints on the interior structure, particularly the central concentration of the planet's mass.  These constraints will inform models of the planet's interior and formation.  This parameter has been measured for bodies in the solar system using radio science observations and for Earth using satellite laser ranging \citep{Rutkowska2010k2}.  Investigations of the tidal dynamics of eclipsing binaries go back as far as the late 19th century \citep{Kreiner2001}. One way to measure \math{k\sb{\rm{2}}} for exoplanets and eclipsing binaries is to detect changes in the occultation timing. For example, \citet{CampoEtal2011apjWASP12b} unsuccessfully attempted to obtain \math{k\sb{\rm{2b}}} for WASP-12b through measurement of the planet's apsidal precession rate using secondary eclipses and radial velocity data.

\section{OBSERVATIONS}

{\em{Spitzer}} observed two secondary eclipses on 9 May 2010 and 8 June 2010 (program ID 60003).  The first observation was performed in the Infrared Array Camera (IRAC) 3.6 {\micron} bandpass (Channel 1) and the second in the IRAC 4.5 {\micron} bandpass (Channel 2).  Both observations were conducted in sub-array mode and consisted of 68,608 frames, lasting 8:09:06.  The Channel 1 observation started at 23:55:32 UTC and ended at 8:04:37 UTC.  The channel 2 observation began at 4:03:15 UTC and ended at 12:12:21.    In an initial version of the Channel 1 data, analyzed with the initial Spitzer reduction pipeline (version S18.18.0), we found a discontinuity in the background level due to separate dark current corrections being applied midway through the observation.  This was subsequently corrected in the {\em Spitzer} archive.

\section{DATA ANALYSIS}
\label{sec:data_analysis}

\subsection{Photometry}
The data were analyzed using the Photometry of Orbits, Eclipses and Transits (POET) pipeline  \citep{StevensonEtal2010Nature,
 CampoEtal2011apjWASP12b, Nymeyer2011, StevensonEtal2012apjHD149026b,
 Stevenson2012GJ436c, Cubillos2013WASP8b, Cubillos2014TrES1}.  We used basic calibrated data (BCD) frames processed through {\em{Spitzer}} pipeline S18.18.0 for the 4.5 {\micron}\ band and S19.1.0 for the 3.6 {\micron}\ data.   Photometry apertures were tested with radii in quarter-pixel increments from 1.5 to 4.0 pixels for the 4.5 {\micron}\ data and 1.75 pixels to 4.0 pixels for the 3.6 {\micron}\ data.   The sky annulus for all apertures had inner and outer radii of 7 and 15 pixels, respectively.

\subsection{Lightcurve Modeling}
POET simultaneously models the combined effects of the eclipse, temporal sensitivity variations (the ``ramp''), and intra-pixel position sensitivity.  POET models these effects separately and combines them in a simultaneous fit, where the measured flux is simply the product of these three models.  The eclipse model follows that of \citet{MandelAgol2002ApJtransits}.  We chose four ramp models for comparison:  no ramp, linear, quadratic, and rising exponential. No other models were considered. POET maps and removes intra-pixel gain variations using the bilinearly-interpolated subpixel sensitivity (BLISS) mapping technique described by \citet{StevensonEtal2012apjHD149026b}.  We removed the first 20\% of the data (13,721 points) because they were taken when the telescope pointing had not yet settled and when the ramp has its greatest impact.  This is the ``preclip'' referred to in Table \ref{table:full_fit}. The pointing settling resulted in a poorly calibrated portion of the BLISS map, which provided little information on intra-pixel gain during the eclipse and thus could not be used to constrain the eclipse model.  

We modeled lightcurves for each aperture size.  At the beginning of each fit, we minimized  \math{\chi\sp2} to find the best fitting set of parameters.  The uncertainties for the data are then rescaled such that the reduced \math{\chi\sp2 = 1}. The same scaling factor is used for all models in a given aperture to allow model comparison (see below); the differences in scaling factors for different models are generally small.  The fit is then re-run.  This procedure accounts for uncertainty over/underestimates in Spitzer's pipeline due to assumptions that do not apply to all analyses.  This approach also accounts for a global average of correlated noise, in the sense that the uncertainties will be larger than the point-to-point scatter if red noise is important.  Uncertainties associated with the best-fit parameters are then found using the differential-evolution Metropolis-Hastings Markov-chain Monte Carlo (MCMC) algorithm \citep{Braak2006DifferentialEvolution}.  We then compared the results across apertures, using the standard deviation of normalized residuals (SDNR) of each lightcurve to find which aperture gives the best fit.  We compared models using the Bayesian Information Criterion (BIC, \citealp{Liddle2008}), which penalizes complex models and overfitting, rewarding simpler models with good fits and fewer parameters.  A smaller BIC values indicates higher probability of a model.  The probability ratio of any two models whose BIC values differ by \math{\Delta \rm{BIC} = \rm{BIC}\sb{2}-\rm{BIC}\sb{1}} is approximately

\begin{equation}\label{eq:deltabic}
\frac{p_2}{p_1} \sim  e\sp{\Delta \rm{BIC}/2}.
\end{equation}

After finding the best aperture size, we varied the BLISS bin size until the BIC values for bilinearly interpolated intrapixel maps and nearest-neighbor-interpolated intra-pixel maps were comparable \citep{StevensonEtal2012apjHD149026b}.

Gaussian priors were placed on the eclipse duration (0.1345 \math{\pm} 0.0017 d) and ingress and egress time  (0.018 \math{\pm} 0.0018 d) for both channels based on the corresponding transit values by \citet{Bakos2009discovery}.  This implicitly assumes that the geometry of the transit is identical to that of the eclipse, an assumption that requires a negligible eccentricity.  For comparison, we calculated the eclipse duration and ingress and egress times from the known system parameters.  These agree with their counterpart transit parameters to within the uncertainty of the transit duration.  Other priors were flat within their bounds.

%

\subsection{Results}

\subsubsection{Channel 1 - 3.6 \math{\mu}m}

\begin{figure}[ht]
\vspace{-5pt}
\includegraphics[width=0.9\linewidth, clip]{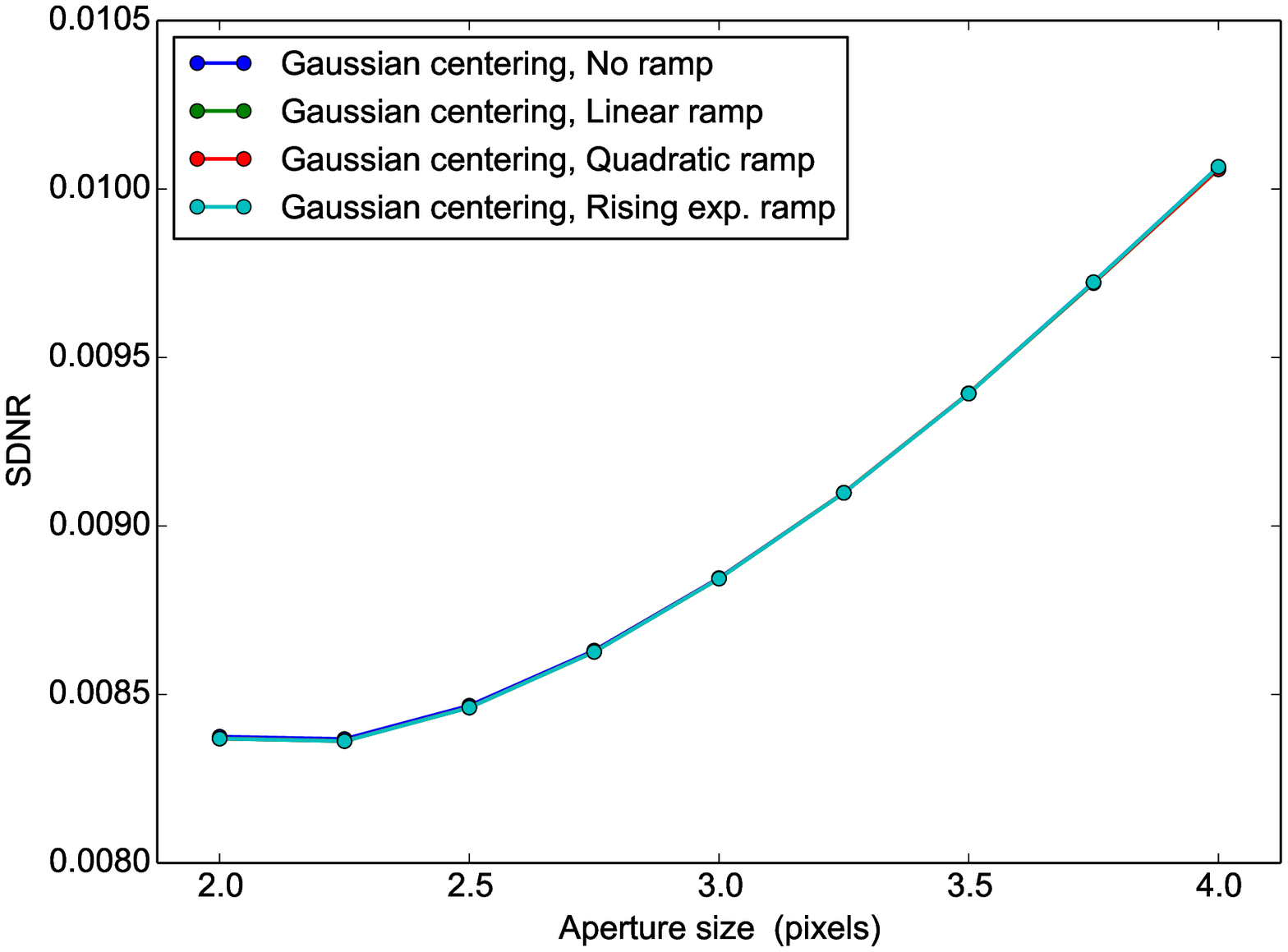}

\figcaption{\label{fig:ch1_apcompare} Comparison of SDNR across all aperture radii and models examined in the 3.6 {\micron} band, showing a minimum at 2.25 pixels.}
\end{figure}

\begin{figure}[ht]
\includegraphics[width=0.9\linewidth, clip]{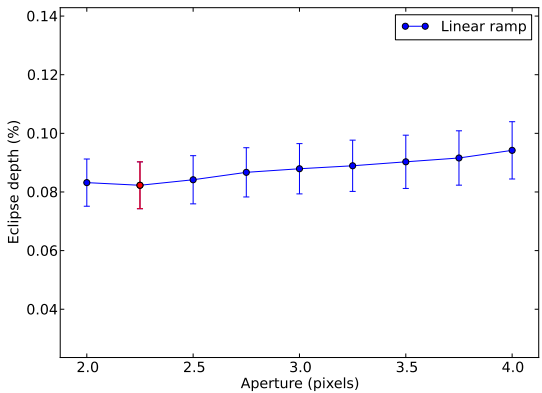}
\figcaption{\label{fig:ch1_depth}
The 3.6 {\micron}  eclipse depth for each aperture.
  The depth corresponding to the aperture radius with the lowest SDNR (2.25 pixels) is highlighted in red. }
\end{figure}

For the 3.6 {\micron} channel, we tested nine apertures ranging between 2.0 and 4.0 pixels in increments of 0.25 pixels.  In each aperture, we tested four temporal sensitivity models:  flat, linear, quadratic, and rising exponential.  Across all four models and nine apertures, the aperture radius yielding the minimum SDNR was 2.25 pixels.  This is shown in Figure \ref{fig:ch1_apcompare}.  After determining the SDNR-optimal aperture, we performed a comparison of the BIC values for each model.    As shown in Table \ref{table:ch1_ramps}, the linear ramp had the lowest BIC value
and therefore best described the data (the BIC value is relative to the lowest one).  The quadratic ramp is the next most plausible model with a probability ratio of \sim180:1, according to Equation \ref{eq:deltabic}.    Figure \ref{fig:ch1_depth} shows a dependency of the eclipse depth on aperture size that would change our eclipse depth by less than 1{\math{\sigma}}.  Figure \ref{fig:joint_rms} shows the RMS of the residuals vs. bin size for these data.

\begin{table}[ht]
\caption{\label{table:ch1_ramps}
Channel 1 Model Comparison at Aperture 2.25 with Preclip}
\atabon\strut\hfill\begin{tabular}{lccc}
    \hline
    \hline
    Ramp Model                 & SDNR       & \math{\Delta} BIC    & Eclipse Depth [\%] \\
    \hline
    Linear                     & 0.0083845  & 0.0  & 0.0823 \\ 
    Quadratic                  & 0.0083846  & 10.4 & 0.0779 \\ 
    Rising Exponential         & 0.0083844  & 11.1 & 0.0835 \\ 
    No Ramp                    & 0.0083909  & 69.3 & 0.0787 \\ 

    \hline
\end{tabular}\hfill\strut\ataboff
\end{table}


\subsubsection{Channel 2 - 4.5 \math{\mu}m}

\begin{figure}[ht]
\vspace{-5pt}
\includegraphics[width=0.9\linewidth, clip]{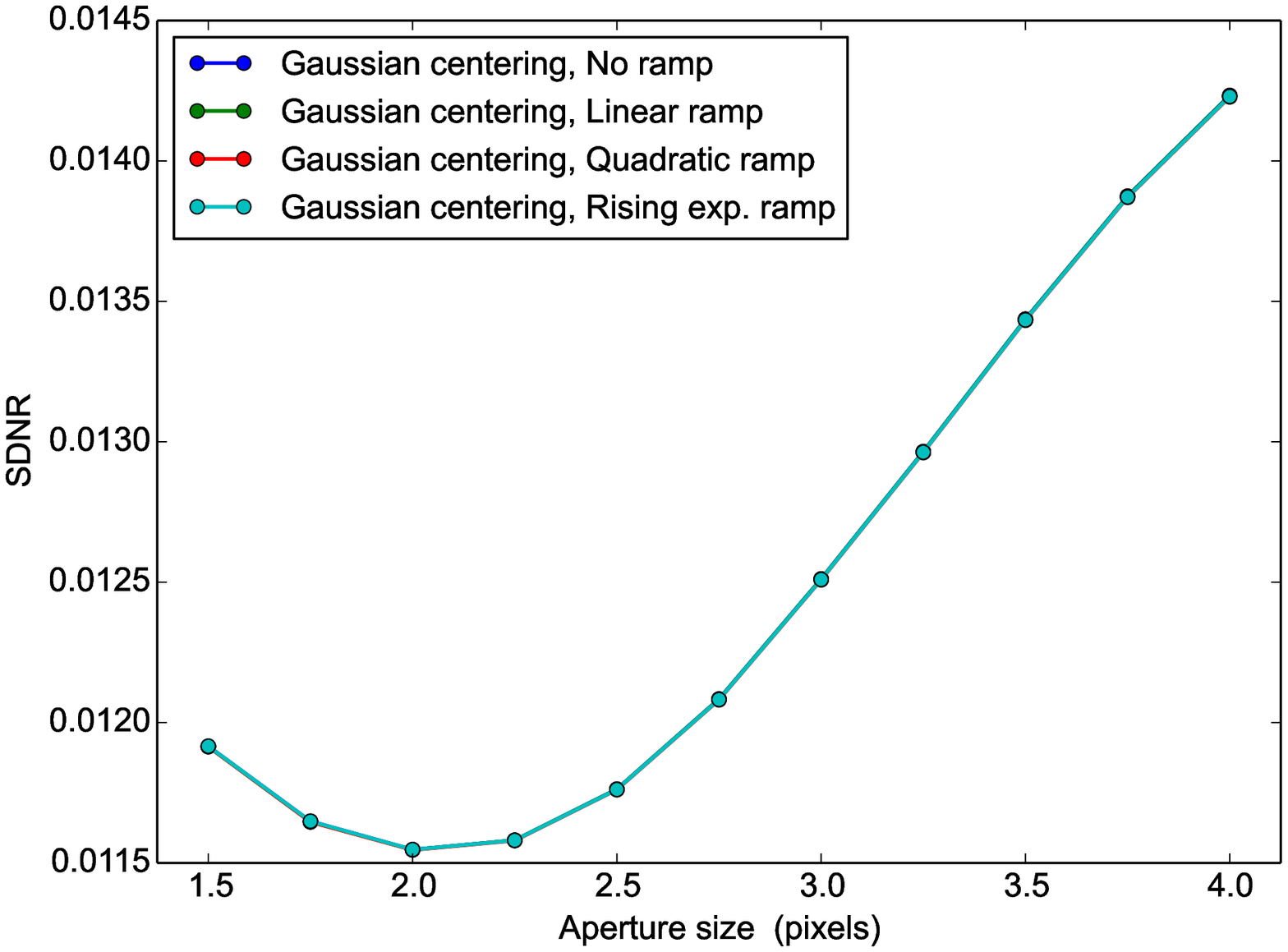}

\figcaption{\label{fig:ch2_apcompare} Comparison of SDNR across all aperture radii and models examined in the 4.5 {\micron} band, showing a minimum at 2.00 pixels.}
\end{figure}

\begin{figure}[ht]
\includegraphics[width=0.9\linewidth, clip]{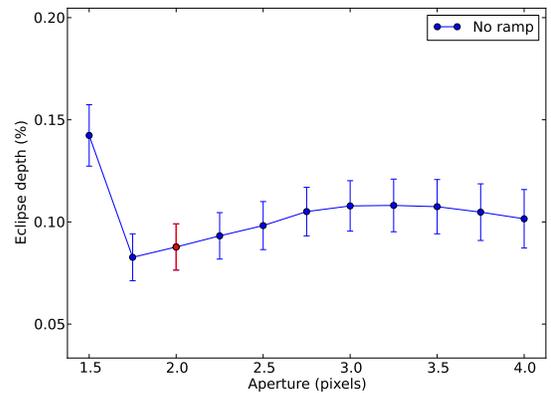}
\figcaption{\label{fig:ch2_depth}
The 4.5 {{\micron} } eclipse depth for each aperture.    The depth corresponding to the aperture radius with the lowest SDNR (2.00 pixels) is highlighted in red.   We note that the extremely small aperture radius of 1.5 pixels produces an anomalously high depth. 
}
\end{figure}

In this channel, we tested 11 apertures ranging between 1.5 and 4.0 pixels in increments of 0.25 pixels.  As with the 3.6 {\micron} channel, in each aperture, we tested four temporal sensitivity models: flat, linear, quadratic, and rising exponential.  Across all four models and 11 apertures, the aperture radius yielding the minimum SDNR was 2.00 pixels, as shown in Figure \ref{fig:ch2_apcompare}.  As shown in Table \ref{table:ch2_ramps}, BIC determined that the model with no temporal variation best describes the data.  The linear ramp is also plausible. Our choice of aperture does not impact the eclipse beyond the 1\math{\sigma} level (Figure \ref{fig:ch2_depth}), except at the implausibly small aperture radius of 1.5 pixels, which produces an anomalously high depth. 

\begin{figure}[ht]
\noindent\centerline{\includegraphics[width=\linewidth, clip]{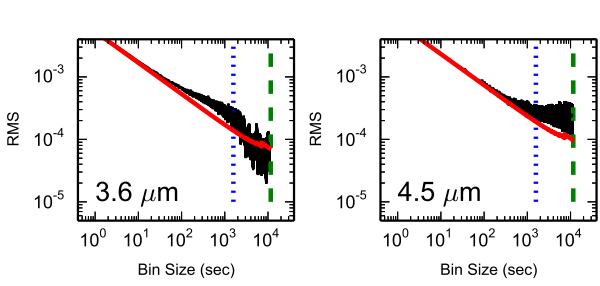}}
\figcaption{\label{fig:joint_rms}
RMS noise {\em vs.} bin size for the 3.6 and 4.5 {\micron} channels in the joint fit. The green and blue lines indicate the eclipse duration and ingress/egress time, respectively.} To assess the red noise at different scales, we calculate the RMS of the residuals
from the best-fit model after averaging neighboring residuals in bins of
increasing sizes.  The vertical black bars indicate the 3\math{\sigma}
range of the RMS estimate at each bin size.  The red line is the
theoretical scaling of the unbinned RMS residual.  At 3.6 {\micron} there is no indication of significant red noise at the eclipse-duration
scale.  See Section 3.3.2 for discussion of the validity of this test
and how it applies to our data. 
\end{figure}

The RMS as a function of bin size for this channel is shown in Figure \ref{fig:joint_rms}.  If the red line is within the black region, there is no significant
detection of red noise at that scale.  The blue and green lines
indicate the eclipse duration and ingress/egress time, respectively.
Note that the black ranges are inherently strongly correlated for this type of calculation.

\citet{Cubillos2016rednoise} has examined this and other red-noise tests,
and finds that the assumptions of this standard test break down at
precisely the most interesting scale: The eclipse duration is roughly
half the length of the data, but the test is only valid for bin sizes
much smaller than the length of the data. 

A standard red-noise adjustment based on this (questionable) test is
to scale the eclipse-depth uncertainties by the ratio of the measured
to theoretical RMS (black to red, in Figure \ref{fig:joint_rms}).  However, our
uncertainties already account for a global average of red noise, being
calculated as the RMS residuals from the best-fit model over the
entire fit.  The standard adjustment is to use uncertainties calculated
from a short run of data chosen from a region that looks like it has
little red noise, or by taking the RMS of \math{1/\sqrt{2}}  times the difference between
adjacent flux values, which assumes that the model changes little
between adjacent points and that small-scale red noise is small.

At 4.5 {\microns} we take our photometric uncertainties to be \math{1/\sqrt{2} } 
of the point-to-point variation RMS, 539.7 {\micro}Jy. We then fit the model to the
data with these uncertainties, compare the measured residual RMS 
to the theoretical RMS at the eclipse scale (Figure \ref{fig:joint_rms}), and scale
the eclipse-depth uncertainty by this ratio, 2.453, resulting in an eclipse
depth of 0.0880\% {\pm} 0.0280\%.

\begin{table}[ht]
\caption{\label{table:ch2_ramps}
Channel 2 Model Comparison at Aperture 2.00 with Preclip}
\atabon\strut\hfill\begin{tabular}{lccc}
    \hline
    \hline
    Ramp Model                 & SDNR       & \math{\Delta} BIC    & Eclipse Depth [\%] \\
    \hline
    No Ramp                    & 0.0115917  & 0.0 & 0.0877 \\ 
    Linear                     & 0.0115912  & 5.8 & 0.0867 \\ 
    Quadratic                  & 0.0115903  & 15.8 & 0.1038 \\ 
    Rising Exponential         & 0.0115909  & 17.4 & 0.0947 \\ 

    \hline
\end{tabular}\hfill\strut\ataboff
\end{table}

\subsubsection{Joint Fit}

\begin{figure*}[ht]
\vspace{-5pt}
\strut\hfill
\includegraphics[width=0.3\textwidth, clip]{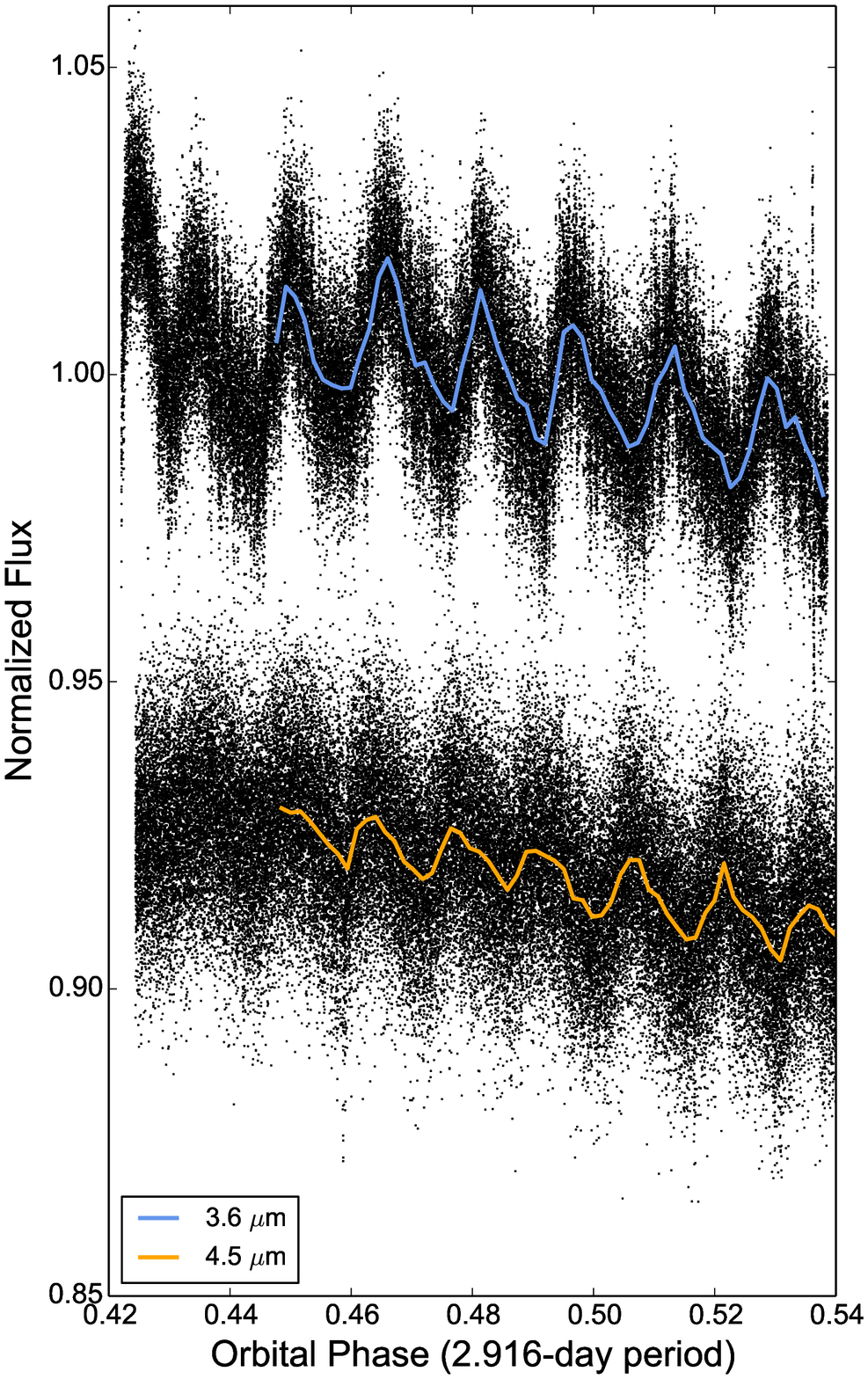}\hfill
\includegraphics[width=0.3\textwidth, clip]{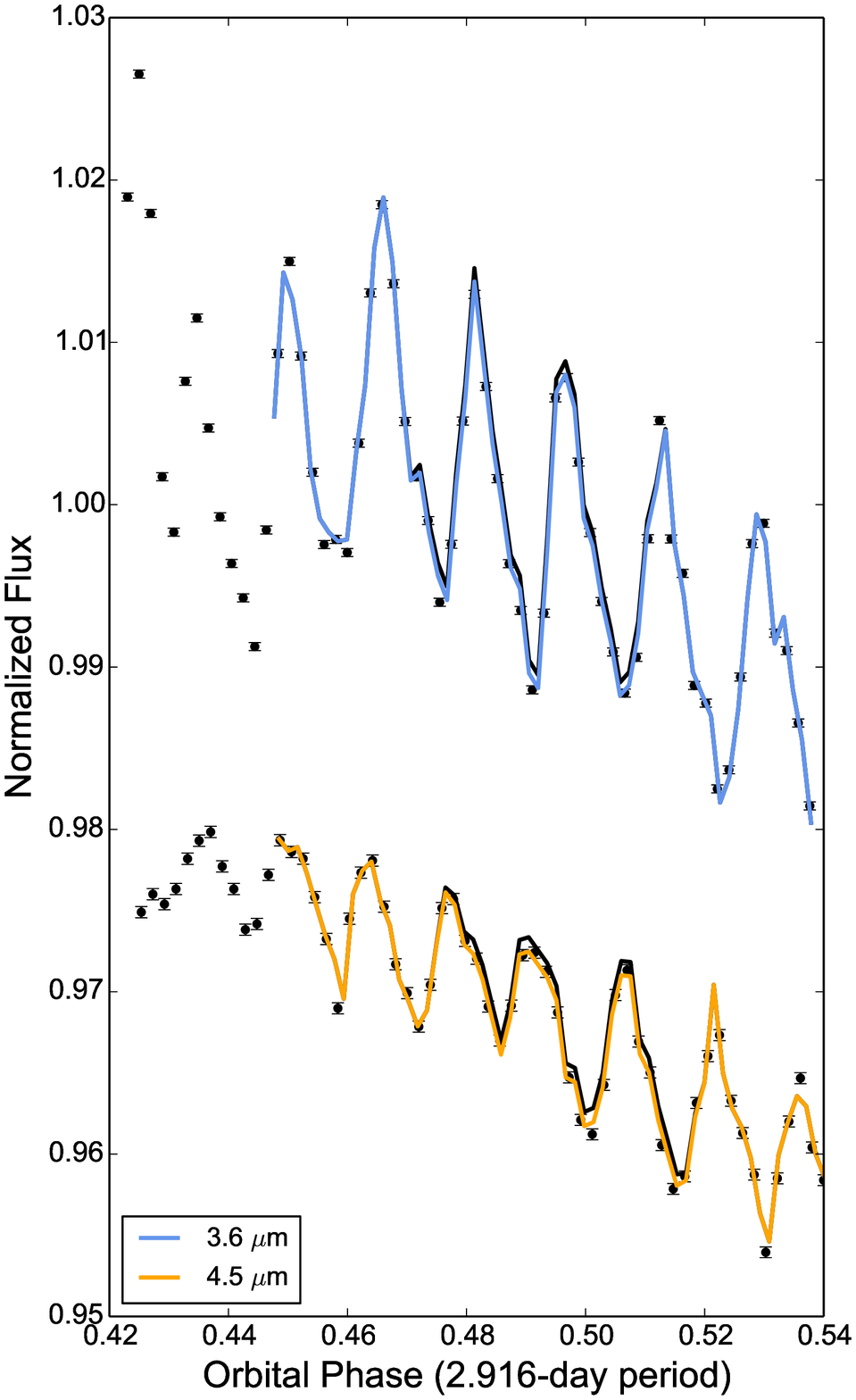}\hfill
\includegraphics[width=0.3\textwidth, clip]{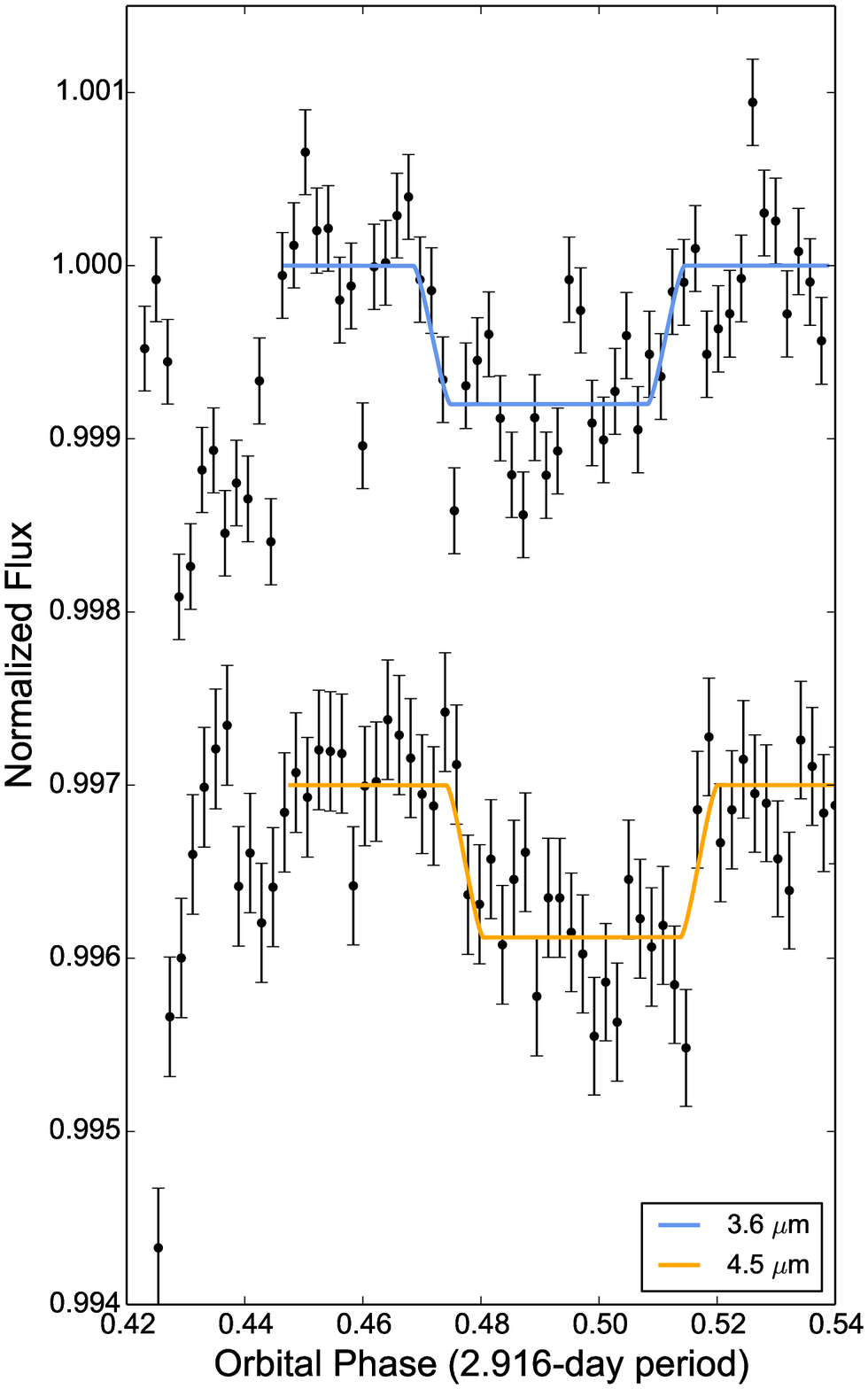}
\hfill\strut
\figcaption{\label{fig:lightcurves}
Secondary-eclipse photometry for HAT-P-13 b in the IRAC  3.6 {\micron} and 4.5 {\micron} bands.  Corresponding best-fit models are represented in orange and blue, respectively.  \textbf{Left:}  Raw photometry with best-fit model showing position- and time-dependent sensitivity correlated with instrument oscillation.  \textbf{Center:}  Binned photometry with the best-fit model (colored lines) and the best-fit model without an eclipse (black lines).  \textbf{Right:}  Binned photometry with time- and position-dependent systematics divided out, leaving only the secondary eclipse and best-fit eclipse models.  Error bars represent 1\math{\sigma} uncertainties.
}
\end{figure*}

The final results we report here come from a simultaneous fit to both eclipses and are summarized in Table \ref{table:full_fit}.  Both eclipses shared eclipse duration and limb crossing time to improve the accuracy of these parameters.  The eclipse midpoints were kept independent to enable future study of their variation in time. The Gelman-Rubin convergence test is satisfied to within 1\%. In pairwise parameter plots, all distributions appear Gaussian with low correlation coefficients.  Individual parameter histograms all appear symmetric and Gaussian.  Parameter trace plots indicate thorough exploration of the full range of each parameter. The final photometry and best-fit models are shown in Figure \ref{fig:lightcurves}. 


\begin{table*}[ht]
\centering
\caption{\label{table:full_fit}
Joint Best-Fit Light Curve Parameters}
\begin{tabular}{r c c}
\hline
\colhead{Parameter}                                           & 3.6 \math{\mu}m       & 4.5 \math{\mu m}     \\
\hline
Array position ($\bar{x}$, pix)                               & 14.25            & 14.10           \\
Array position ($\bar{y}$, pix)                               & 14.88            & 14.92           \\
Position consistency\tablenotemark{a} ($\delta\sb{x}$, pix)   & 0.008            & 0.012           \\
Position consistency\tablenotemark{a} ($\delta\sb{y}$, pix)   & 0.008            & 0.010           \\
Aperture size (pix)                                           & 2.25             & 2.00            \\
Sky annulus inner radius (pix)                                & 7.0              & 7.0             \\
Sky annulus outer radius (pix)                                & 15.0             & 15.0            \\
Eclipse depth (\%)                                            & 0.0801(81)       & 0.088(28)       \\
Brightness Temperature (K)                                    & 1732(75)         & 1573(93)        \\
Midpoint (orbits)                                             & 0.49154(80)      & 0.49711(83)     \\
Eclipse midpoint (BJD\sb{UTC}-2,450,000)                   & 5326.7020(23)  & 5355.8807(24) \\
Eclipse midpoint (BJD\sb{TDB}-2,450,000)                      & 5326.7027(23)  & 5355.8815(24) \\
Eclipse duration (hr)                                         & 3.215(28)        & 3.215(28)       \\
Ingress/Egress time (hr)                                      & 0.439(28)        & 0.439(28)       \\
System flux: $F\sb{s}$ (\micro Jy)                            & 74635.7(3.5)     & 46330.3(2.8)    \\
Ramp: $R(t)$                                                  & Linear           & None            \\
Ramp linear term                                              & 0.0183(21)       & $\cdots$        \\
BLISS map ($M(x,y)$)                                          & Yes              & Yes             \\
Minimum number of points per bin                              & 4                & 4               \\
Total frames                                                  & 68608            & 68608           \\
Frames used                                                   & 53250            & 54386           \\
Rejected frames (\%)                                          & 1.48            & 0.09           \\
Preclip frames (\%)                                          & 20            & 20          \\
Free parameters                                               & 6                & 3               \\
AIC value                                                     & 107644.9         & 107644.9        \\
BIC value                                                     & 107731.1         & 107731.1        \\
SDNR                                                          & 0.0083616        & 0.0115475       \\
Uncertainty scaling factor                                    & 0.951            & 0.531           \\
\hline\end{tabular}
\begin{minipage}[t]{0.65\linewidth}
\tablenotetext{1}{RMS frame-to-frame position difference.}
\end{minipage}
\end{table*}

\section{Atmosphere}

HAT-P-13 b is expected to have an equilibrium temperature of 1626 \pm 42 K based on its proximity to its host star and an assumed Bond albedo of zero and efficient energy redistribution.  Its scale height for a solar-abundance hydrogen-helium composition is expected to be ~500 km.  The predicted adiabatic lapse rate is ~1.5 K km\sp{-1}.

Our secondary eclipse observations enabled modeling of the dayside atmosphere of HAT-P-13 b.  The 3.6 {\micron} band is particularly sensitive to the spectral signature of methane (CH\sb{4}) while the 4.5 {\micron} band captures spectral features associated with CO and CO\sb{2}.  Whether these chemical species show up as emission or absorption signatures depends on whether temperature increases or decreases with altitude at
the depths we are probing.  This allows us to model the pressure-temperature, \math{T(p)}, profile of HAT-P-13 b, albeit with some degeneracy. 

To model the dayside spectrum and atmosphere of HAT-P-13 b, we use the method described by \citet{MadhuSeager2010, MadhuSeager2009}, with the addition of Markov-chain Monte Carlo and Bayesian analysis as described in the supplementary information of \citet{StevensonEtal2010Nature}.  This model performs one-dimensional, line-by-line, radiative-transfer calculations for a plane-parallel atmosphere, assuming global energy balance, hydrostatic equilibrium, and local thermodynamic equilibrium.  The spectrum of the star is interpolated from a grid of Kurucz models \citep{Kurucz2004}. Having fewer spectral samples of this planet's atmosphere (2) than free model parameters (10), it is impossible to determine a unique model of the planet's atmosphere.  However, certain chemical configurations can be ruled out by modeling deviations from a solar-abundance spectrum.   Past \textit{Spitzer} observations of other planets have indicated methane deficiency \citep{StevensonEtal2010Nature} and an unexpectedly high C--O ratio \citep{MadhuNature, Stevenson2014, Line2012}. 

\begin{figure}[ht]
\vspace{-5pt}
\includegraphics[width=0.9\linewidth, clip]{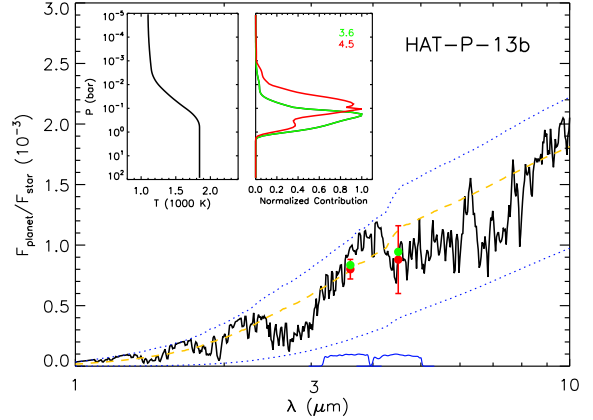}
\figcaption{\label{fig:atmosphere}
Well-fitting model spectrum for HAT-P-13 b with the Spitzer bandpasses (main) and the corresponding pressure-temperature profile (left inset) and normalized contribution functions (right inset).  The dashed yellow line is the corresponding blackbody spectrum for a 1700 K body and the blue curves show other the blackbody spectra for the highest and lowest  temperatures (1843 K and 1100 K, respectively) in the pressure-temperature profile depicted in the inset. Note the logarithmic scale of the abscissa, with each tick denoting a 1 {\micron} interval.
}
\end{figure}

From modeling the atmosphere of HAT-P-13 b, we find that it has efficient day-night energy redistribution. Our data are consistent with a blackbody and  with a wide range of compositions, including solar.  Given the large uncertainty on the 4.5 {\micron} eclipse depth, our data can be explained by a wide range of temperature profiles. For the altitudes probed (Figure \ref{fig:atmosphere}, right inset), an isothermal atmosphere at 1700 K would yield a blackbody spectrum consistent with the data (yellow dashed line in Figure \ref{fig:atmosphere}). Alternately, the data are also consistent with a fiducial solar-abundance composition with no thermal inversion (black line in Figure \ref{fig:atmosphere} and its left inset). Finally, since the brightness temperature in the 4.5 {\micron} channel, within the 1\math{\sigma} uncertainties, can be slightly higher than that in the 3.6 {\micron} channel, we cannot robustly rule out a thermal inversion. However, a strong thermal inversion is still unlikely. Future observations in the near-infrared, e.g., with the {\em Hubble Space Telescope} Wide Field Camera 3 (1.1-1.7 {\micron}) or from the ground in the J, H, and K bands, could provide more robust constraints on the atmospheric temperature profile \boldmath{T(p)} and composition of this planet. The uncertainty at 4.5 {\micron} was increased to account for red noise, as presented in Figure \ref{fig:joint_rms}. As discussed in Section 3.3.2, this test has questionable validity.  The original unscaled uncertainty of 0.011\% ruled out a blackbody and a thermal inversion.


\section{Orbit and Interior Structure}
\subsection{Orbit}

The phases of the observed secondary eclipses provide additional constraints on the orbit of HAT-P-13 b.  The 3.6 {\micron} event occurs at a phase of 0.49154 \math{\pm} 0.00080 and the 4.5 {\micron} event occurs at a phase of 0.49711 \math{\pm} 0.00083. We note that the phases are widely separated, with the 4.5 {\micron} eclipse occurring 23 \math{\pm} 4 minutes later than the 3.6 {\micron} eclipse.  One possibility is that the difference is a result of differential brightness across the planet's surface in each of these bands. However, the timing difference is comparable to the best-fit ingress and egress time of the eclipses (26 \math{\pm} 2 minutes), which carries the implausible implication of the brightness centers in the two bands being on opposite sides of the planet.  Secular apsidal precession is also too small to explain this difference over a one-month timescale.

As a preliminary estimate, we can assess the impact of our eclipse-phase measurements on the solution through comparison with orbital parameters from prior work. After a first-order light-time correction of 42.5 s, these eclipse phases respectively indicate \math{e \cos \omega} values of -0.0135 \math{\pm} 0.0013 and -0.0048 \math{\pm} 0.0013. These are in rough agreement with the radial-velocity value of \math{e \cos \omega=-0.0099 \pm 0.0036} \citep{Winn2010RV}. To predict the effect of our \math{e \cos \omega} observation on the orbital solution before joint modeling, we took their weighted means individually with the \citet{Winn2010RV} data, yielding -0.0132 \math{\pm} 0.0012 and -0.0054 \math{\pm} 0.0012, respectively. Substituting these averages and combining them with the original \citet{Winn2010RV} \math{e \sin \omega} value of 0.0060 \pm 0.0069, we find values of \math{e = 0.0145 \pm 0.0031} and \math{\omega = (155 \pm 12)\sp{\rm{o}}} for the 3.6 {\micron} event and \math{e = 0.0081 \pm 0.0052} and \math{\omega = (132 \pm 37)\sp{\rm{o}}} for the 4.5 {\micron} event. Because of the \math{5\sigma} separation between these two events, we elect not to take their weighted average. The calculations instead tune our expectations for more sophisticated joint modeling.

In our detailed MCMC model, we separately incorporated both secondary-eclipse midpoint times into a joint fit including radial-velocity (RV) data (see below), transit photometry \citep{Southworth2012transits}, and amateur \citep{Poddany2010amateur} and professional transit midpoints.  We use HIRES RV data tabulated by \citet{Knutson2014}, which include all previous measurements by \citet{Winn2010RV} and \citet{Bakos2009discovery}. One outlier radial velocity measurement in the \citet{Knutson2014} dataset at BJD 2455945 was discarded.  None of these data were taken during a transit, so the Rossiter-McLaughlin effect was not modeled.  All timestamps on the data were converted to a consistent BJD (TDB) time standard as prescribed by \citet{Eastman2010time}. The mid-transit times and their origins are all listed in Table \ref{tab:ttv}.  Only ETD transit data with a quality rating of 1--3 were used. The limb-darkening parameters were fixed to interpolated tabulated values found by \citet{Winn2010RV}. 

The model we use for the orbit fit is briefly described in \citet{Stevenson2012GJ436c}.  The model uses 13 free parameters, and follows the formalism of \citet{Gillon2009}.  The code, \textit{Photometry, Event Timing, and Radial-velocity Analysis}, PETRA, minimizes \math{\chi\sp{2}} for the aforementioned data using the transit photometry models of \citet{MandelAgol2002ApJtransits}, radial velocity models described by \citet{Paddock1913}, and a linear ephemeris for transit and eclipse timing data with provisions for apsidal motion as described by \citet{CampoEtal2011apjWASP12b}.  The parameters for planet b are the reduced transit impact parameter (\math{b\sp{\prime} = b\frac{1+e\sin{\omega}}{1-e^2}}); the scaled semi-amplitude (\math{K_2 = K\sqrt{1-e^2}P^{1/3}}); the planet-star radius ratio; the transit duration; the transit midpoint time \math{T\sb{0b}}; the period \math{P\sb{b}}; and the Laplace vector components \math{e\sb{b} \cos \omega\sb{b}} and \math{e\sb{b} \sin \omega\sb{b}}.  Planet c's parameters are its scaled semi-amplitude; Laplace vector components; predicted transit midpoint, and orbital period.  The fit also models the velocity and secular acceleration of the star HAT-P-13. The model uses the stellar density calibration of \citet{Enoch2010density} to take these modeling parameters and determine the physical parameters of the system.   We estimate photometric and radial velocity jitter (0.25\% of total stellar flux and 4.2 m s\math{^{-1}}, respectively) from an initial best fit to all available data and add these jitter estimates to the instrumental error bars, accordingly.

\begin{figure}[ht]
\vspace{-5pt}
\includegraphics[width=0.9\linewidth, clip]{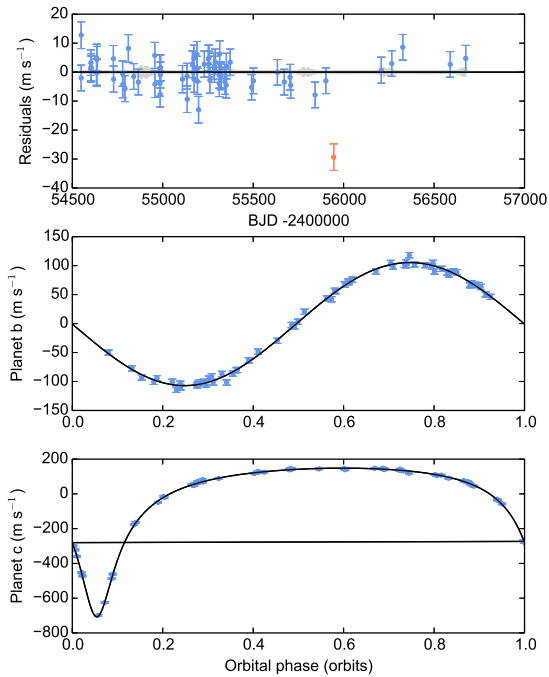}
\figcaption{\label{fig:orbit_rv}
Residuals of included radial velocity data (top, blue points) against the best-fitting orbit model. The middle and bottom plots show the best-fitting radial-velocity predictions (black lines) for each planet shown with folded RV data. The shaded gray region around the residual zero line shows the 1\math{\sigma} model prediction uncertainty (not the per-point uncertainty). The single excluded point (orange) does not dramatically alter the best-fit parameters when included, but increases the estimated jitter by \sim20\%.}
\end{figure}

\begin{figure}[ht]
\vspace{-5pt}
\includegraphics[width=0.9\linewidth, clip]{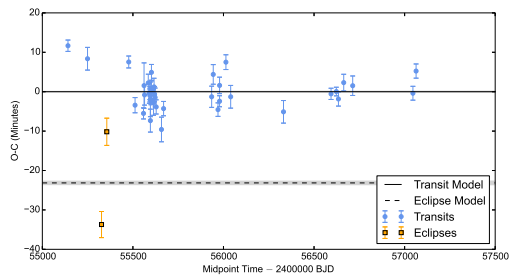}
\figcaption{\label{fig:orbit_oc}
Timing residual (observed minus calculated, O-C) plots for transit and eclipse midpoint timing data.  This figure shows the residuals of timing data relative to predictions for a circular orbit. The best-fitting models for transit and eclipse data are shown by solid and dashed lines, respectively. The shaded gray regions around each model show the 1\math{\sigma} model prediction uncertainty.}
\end{figure}

We ran a Metropolis-Hastings Markov chain for one million iterations to explore the parameter space and estimate uncertainties.  Pairwise marginalization of the parameters shows low correlations, except between three of the transit parameters and the period and transit midpoint time.  Priors were uniform and bounded. Inspection of trace plots shows good exploration within the bounds of each parameter.  Posterior histograms of the individual parameters are roughly Gaussian and descend to zero within the boundaries.  Where necessary, we estimate 1\math{\sigma} confidence intervals by finding a 68\% interval centered on the mode of each posterior distribution.  For Rayleigh-distributed parameters like eccentricity, the median value tends to be larger than the mode, leading to asymmetric uncertainty estimates.  Table \ref{table:orbit} summarizes our fit results. Figure \ref{fig:orbit_rv} shows the best-fitting  radial velocity model and its residuals. Figure \ref{fig:orbit_oc} shows the O-C (residual timing) plot for transit and eclipse data.  We find that the best-fitting eccentricity of planet b's orbit is \math{0.0093\sp{+0.0044}\sb{-0.0016}}, and \math{\omega = (202 \sp{+12}\sb{-46})\degree}.  Because the eccentricity is the Euclidean norm of two normally distributed components, its distribution has a long tail and a median value of 0.011, higher than the mode. The difference in \math{\omega} between planets b and c is \math{\Delta\omega = (27\sp{+14}\sb{-46})\sp{\rm{o}}}, not inconsistent with apsidal alignment.

To further test whether the apsides are aligned, we compute the ratio of the likelihood of aligned apsides to unaligned apsides, otherwise known as the Savage-Dickey density ratio, \citep{Verdinelli1995}.  We use the normalized histograms of the distribution of the difference between the two \math{\omega} values, \math{\Delta \omega = \omega\sb{b} - \omega\sb{c}}, obtained from the MCMC run to estimate a probability density function.  We compare this probability density where \math{\Delta \omega = 0} and to the prior probability density given that the two apsides are randomly oriented according to a uniform distribution (\math{1/ 2\pi}).  This test shows that the aligned model is 0.50 times as likely as the unaligned model, so the planets are about as likely to have their apsides aligned as not.  The uncertainty associated with \math{e_b\sin{\omega_b}}---nearly an order of magnitude larger than that of \math{e_c\sin{\omega_c}}---is a key source of this ambiguity, and can be reduced with continued radial velocity measurements.

\subsection{Interior}

Confirmation of apsidal alignment and measurements of the eccentricity of the inner planet enable study of the planet's interior. \citet{BBL2009k2} outline a method for recovering planet b's Love number, \math{k\sb{2b}}.  This is further explored by \citet{Kramm2012theory} and \citet{BeckerBatygin2013}. For planets describable as fluids, the more centrally condensed the mass of the planet, the less of its outer mass responds to tides and the lower the value of \math{k\sb{2}}.  Planets with more uniform density distributions are more responsive to tides and thus have a higher value of \math{k\sb{2}}.  Planets with some mechanical rigidity tend to have lower values.  The value of \math{k\sb{2}} ranges from 0 for highly condensed planets) to 1.5 for uniform-density spheres \citep{RagozzineWolf2009interiors}.  For Earth, \math{k\sb{2}} has been measured relatively precisely through satellite laser ranging to be 0.3011 \citep{Rutkowska2010k2}. Larger and less rigid planets like Jupiter and Saturn have values of 0.49 and 0.32, respectively \citep{RagozzineWolf2009interiors}.  

\begin{figure*}[ht]
\vspace{-5pt}
\strut\hfill
\includegraphics[width=0.9\linewidth, clip]{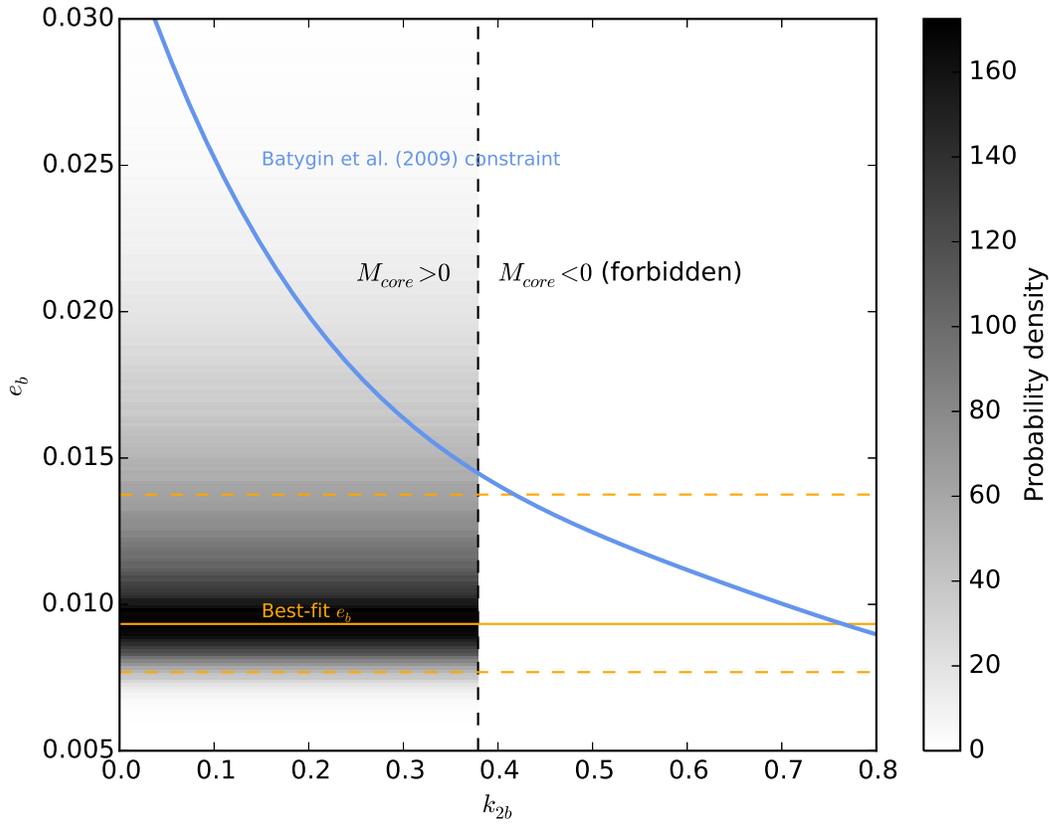}
\hfill\strut
\figcaption{\label{fig:ek2}
Our best-fit eccentricity, with a 1\math{\sigma} confidence interval (solid and dashed orange lines, respectively), and its MCMC-generated posterior probability density distribution (black shading). The blue line is the model relating \math{k\sb{2b}} to the eccentricity of \citet{BBL2009k2} and the upper limit of \math{k\sb{2b}} corresponding to models of HAT-P-13 b with zero core mass derived by \citeauthor{Kramm2012theory} (\citeyear{Kramm2012theory}, vertical dashed line, \math{k\sb{2b} < 0.379}).  
}
\end{figure*}

\citet{BBL2009k2} provide a fourth-order polynomial relating the measured orbital eccentricity of planet b to \math{k\sb{2b}}.  This is further confirmed by \citet{Kramm2012theory} using the observations of \citet{Winn2010RV} to limit \math{k\sb{2b}} to 0.265 -- 0.379.   The \citet{Kramm2012theory} models have two layers:  one containing an envelope of hydrogen, helium and metals; and a rocky core.  Values of \math{k\sb{2b}} above the upper limit found by \citet{Kramm2012theory} correspond to interior models with no rocky core.  Our observations do not rule out apsidal alignment.  Under the same assumptions, 75\% of our MCMC posterior distribution of \math{e\sb{b}} falls below the critical eccentricity of 0.0145, below which modeled core mass vanishes.  The 95\% upper limit for the eccentricity is 0.021.  This distribution is illustrated in Figure \ref{fig:ek2}.

This constraint on \math{k\sb{2b}} is calculated from the best-fit eccentricity from the MCMC run, the distribution of which is skewed by the relatively large uncertainty in \math{e\sb{b} \sin \omega\sb{b}}.  It also depends on the assumption of apsidal alignment.  Only about 2\% of the values of \math{\omega\sb{b}} fall into the full range of values of \math{\omega\sb{c}} in the Markov chain, which spans 1.5\math{\degree}.  The distribution of the aligned subset of eccentricity values is narrower and more symmetric, with a mean of 0.0088 \math{\pm} 0.0009.  Under the relation of \citet{BBL2009k2}, this corresponds to \math{k\sb{2b} = 0.81 \pm 0.10}.  Since \citet{Kramm2012theory} find that values above 0.379 correspond to vanishing core mass, this result may indicate that HAT-P-13 b has a small or nonexistent rocky core. This unexpectedly high estimate is below the theoretical maximum value of \math{k\sb{2b}, 1.5}, which corresponds to a uniform-density sphere. Because the interior structure of planets must obey both hydrostatic equilibrium and the equation of state of its component matter, a true uniform-density sphere of this size is highly implausible. For an \math{n=1} polytrope with a vanishing core, the maximum value of \math{k\sb{2b}} is \sim 0.52 \citep{Storch2015}. Our estimate is within 3\math{\sigma} of this limit.

An alternative interpretation can stem from forcing the assumption that \math{\omega\sb{b}= \omega\sb{c}}, where \math{\omega\sb{c}} is comparatively well known.  From our eclipse-derived measurement of \math{e\sb{b} \cos \omega\sb{b} = -0.00946 \pm 0.00088}, we find that \math{e\sb{b} = 0.0095 \pm 0.0009} and the majority of \math{k\sb{2b}} values are larger than 0.379.  Again, this result suggests a very low core mass.

\begin{table*}[ht]
\centering
\caption{\label{table:orbit}
Joint Orbital Fit MCMC Parameters}
\begin{tabular}{rr@{\,{\pm}\,}lr@{\,{\pm}\,}lr@{\,{\pm}\,}l}
\hline
\hline
\colhead{MCMC Parameters}                                              &  \mctc{} \\
\hline
Planet-to-star radius ratio			                             &   \mctc{      0.0860 $\pm$  0.001    }		\\
Scaled impact parameter				                             &   \mctc{      0.729  $\pm$ 0.030    }		\\
Transit duration	(d)											 &   \mctc{      0.13942   $\pm$   0.0024}		\\
Transit midpoint	(BJD\sb{TDB})											 &   \mctc{      2454779.93106 $\pm$     0.00030 }	\\
Planet b orbital period       (d)                                &   \mctc{       2.91624039   $\pm $   0.00000081}	\\
Linear limb-darkening term                                       &   \mctc{      0.306800        }			\\
Quadratic limb-darkening term                                    &   \mctc{      0.325100        }			\\
\math{e\sb{\rm{b}} \sin \omega\sb{\rm{b}}}                       &   \mctc{       -0.0035 $\pm$ 0.0080   }\\
\math{e\sb{\rm{b}} \cos \omega\sb{\rm{b}}}                       &   \mctc{       -0.0087 $\pm$ 0.0009   }\\
Scaled planet b semi-amplitude   (m s\sp{-1} d\sp{1/3} )         &   \mctc{      151.7 $\pm$ 1.0    }\\
System RV offset         (m s\sp{-1})                            &   \mctc{       -67.0 $\pm$ 0.8}      \\
RV acceleration term     (m s\sp{-1}a\sp{-1})                    &   \mctc{      19.3$\pm$ 0.4    }\\
Planet c transit midpoint   (BJD\sb{TDB})                        &   \mctc{   2455757.69 $\pm$0.19 }  \\
Planet c orbital period     (d)			                         &   \mctc{   445.81 $\pm$ 0.10      }\\
Planet c scaled semi-amplitude	 	(m s\sp{-1} d\sp{1/3} )		         &   \mctc{       2466.4  $\pm$ 6.4  }\\
\math{e\sb{\rm{c}}\sin \omega\sb{\rm{c}}}                        &   \mctc{       0.0528 $\pm$ 0.0023     }\\
\math{e\sb{\rm{c}} \cos \omega\sb{\rm{c}}}                       &   \mctc{       -0.6536 $\pm$ 0.0020    }\\
\math{\chi\sp{2}}                                        &   \mctc{       3972.9      }   \\
Degrees of Freedom                                        &   \mctc{       3810     }   \\
\math{\chi\sp{2}\sb{rd}}                                        &   \mctc{       1.04      }   \\

Number of Data Points in Fit                                     &   \mctc{       3825       }   \\
BIC                                                        &   \mctc{  4096.6   }   \\
\hline 
\colhead{Derived Parameters}\\
\hline
\math{e\sb{b}} 		& \mctc{$0.0093	\sp{+0.0044}\sb{	-0.0016}$}\\
\math{\omega\sb{b} (\sp{\circ})} & \mctc{$202	\sp{+12}\sb{	-46}$} \\
Planet b semi-amplitude (m s\sp{-1}) & 	\mctc{$106.2	\sp{+0.3}	\sb{-1.2}$}\\
Transit impact parameter 	&	\mctc{$0.732\sp{+0.034}\sb{-0.020}$}\\
Planet b semi-major axis (AU) & \mctc{$0.04313\sp{+0.00033}\sb{-0.00026}$}\\
Inclination (degrees) 	&	\mctc{$82.2\sp{+0.6}\sb{-0.8}$}\\
Stellar mass (solar masses) 	&	\mctc{$1.261\sp{+0.029}\sb{-0.023}$}\\
Stellar radius (solar radii) 	&	\mctc{$1.73\sp{+0.10}\sb{-0.09}$}\\
Stellar density (solar density) &		\mctc{$0.243\sp{+0.02}\sb{-0.04}$}\\
\math{a_b/R\sb{*}} 	&	\mctc{$5.36\sp{+0.21}\sb{-0.30}$}\\
\hline\end{tabular}
\begin{minipage}[t]{0.65\linewidth}

\end{minipage}
\end{table*}

\begin{table}[ht]
\centering
\caption{\label{tab:ttv} Transit Timing Data.}
\begin{tabular}{lllc@{ }}
\hline
\hline
Mid-Transit Time  &  Uncertainty          &   Source  \\
 (BJD\sb{TDB})    &                       &                             \\
\hline
2457063.35092	&0.00125	&Bretton\tablenotemark{a} \\
2457045.84957	&0.00122	&Benni\tablenotemark{a} \\
2456713.39949	&0.00172	&Dittler\tablenotemark{a} \\
2456663.82395	&0.00147	&Garlitz\tablenotemark{a} \\
2456634.65866	&0.00125	&Naves\tablenotemark{a} \\
2456622.99501	&0.00076	&Shadick\tablenotemark{a} \\
2456593.8322	&0.00099	&Benni\tablenotemark{a} \\
2456331.36740	&0.00199	&Gaitan\tablenotemark{a} \\
2456039.74601	&0.002		&Shadick\tablenotemark{a} \\
2456013.50595	&0.0013		&Carre\tablenotemark{a} \\
2455978.50697	&0.00145	&Poddan\'{y}\tablenotemark{a} \\
2455978.50418	&0.00097	&Garcia\tablenotemark{a}  \\
2455969.75398	&0.00119	&Shadick\tablenotemark{a}  \\
2455943.51402	&0.00173	&Gonzalez\tablenotemark{a}  \\
2455934.76136	&0.00187	&Emering\tablenotemark{a}  \\
2455657.71276	&0.00217	&Shadick\tablenotemark{a} \\
2455628.55433	&0.00125	&Sergison\tablenotemark{a} \\
2455602.31426	&0.00139	&Poddan\'{y}\tablenotemark{a} \\
2455584.81496	&0.00153	&Dvorak\tablenotemark{a} \\
2455511.90502	&0.00134	&Shadick\tablenotemark{a} \\
2455476.91775	&0.00105	&Shadick\tablenotemark{a} \\
2455613.97390	&0.00225	&\citet{Fulton2011}\\
2455616.89290	&0.00152	&\citet{Fulton2011}\\
2455619.80786	&0.00134	&\citet{Fulton2011}\\
2455622.72351	&0.00166	&\citet{Fulton2011}\\
2455564.39916	&0.0018		&\citet{Nascimbeni2011}\\
2455593.56187	&0.00115	&\citet{Nascimbeni2011}\\
2455596.47702	&0.00305	&\citet{Nascimbeni2011}\\
2455599.39307	&0.00076	&\citet{Nascimbeni2011}\\
2455602.31108	&0.00167	&\citet{Nascimbeni2011}\\
2455558.56342	&0.00098	&\citet{Pal2011}\\
2455561.48456	&0.004		&\citet{Pal2011}\\
2455590.64563	&0.00179	&\citet{Pal2011}\\
2455141.55297	&0.001		&\citet{Szabo2010}\\
2455249.45157	&0.002		&\citet{Szabo2010}\\
2455593.56147	&0.00115	&\citet{Southworth2012transits}\\
2455596.47327	&0.00202	&\citet{Southworth2012transits}\\
2455599.39446	&0.001		&\citet{Southworth2012transits}\\
2455669.38140	&0.00126	&\citet{Southworth2012transits}\\

\hline
\end{tabular}
\begin{minipage}[t]{0.95\linewidth}
\tablenotetext{1}{The Transiting ExoplanetS and Candidates group (TRESCA, http://var2.astro.cz/EN/tresca/index.php) supply their data to the Exoplanet Transit Database (ETD), http://var2.astro.cz/ETD/) which performs the uniform transit analysis described by \citet{Poddany2010amateur}.  The ETD web site provided the numbers in this table, which were converted from HJD (UTC) to BJD (TDB).}
\end{minipage}
\end{table}

\section{Conclusions}

Our observations of HAT-P-13 b's secondary eclipse timing have allowed us to refine its eccentricity and inform models of its interior structure better than with radial velocity and transit observations alone.  The pressure-temperature profile from our atmospheric model fits provides a constraint on the envelope density of HAT-P-13 b, which can also inform interior structure models and further be used to constrain \math{k\sb{2b}}. Our lowered best-fit eccentricity suggests a higher range of \math{k\sb{2b}} values and low probability of HAT-P-13 b having a substantial core mass.   

This conclusion rests on  the assumption that all libration between planets b and c has damped out and the apsides are aligned.  While our data are not inconsistent with apsidal alignment, the uncertainties of the measurements of \math{\Delta \omega} do allow separation of tens of degrees.  Bayesian comparison of the aligned and non-aligned hypotheses is inconclusive. 

If we force apsidal alignment, the measured eccentricity and its uncertainty become consistent with the value of \math{e_b \cos\omega_b} measured with eclipse phases.  This eccentricity value (\sim0.009) is consistent with the best-fit value and consistent with very high values of \math{k\sb{2b}}.  

If the conditions for the measurement of \math{k\sb{2b}} are indeed satisfied and the measurement of the eccentricity can be taken at face value, then the implausibility of a low core mass may require adjustment of the interior models.

The less tantalizing possibility is that the central assumption is invalid: HAT-P-13 b and c may simply not be apsidally aligned and their mutual libration has not damped out.  In this case, this system may be an inappropriate candidate for probing exoplanet interiors in this manner. Our measurement of the eccentricities and arguments of periapsis for both planets may not represent the fixed point of libration in the phase space of \math{e\sb{b}} and \math{\Delta \omega} necessary to infer \math{k\sb{2b}}. If this is the case, then our measurement of these parameters represents a single snapshot of the evolution of these parameters about the fixed point. As the librational period of this system is \math{\sim 10^5} years \citep{Mardling2007}, it is unlikely that this evolution can be observed within a human lifetime with present measurement precision. While the fixed-point eccentricity may be close to our measured value (e.g., within \math{\sim}0.01), the sensitivity of \math{k\sb{2b}} to errors in \math{e\sb{b}} is too high for us to produce any  meaningful estimates of \math{k\sb{2b}} for the non-aligned case.

One of the major caveats of our timing measurements is the significant difference in secondary eclipse times between channels.   The two eclipse midpoints differ by 23 \math{\pm} 4 minutes, which is approximately the same as HAT-P-13 b's limb crossing time.  If this difference is due to wavelength-dependent asymmetric brightness distributions, it requires hot spots to be extremely far from the sub-stellar point and in opposite directions for the two channels.  One possible explanation for the difference is correlated noise.  Our procedure for estimating uncertainties already considers a global average of correlated noise.  Secular apsidal precession is too small to explain this difference over the month separating these observations.  Eclipse-timing variations on the scale of the transit-timing variations may be possible, but the difference between the eclipse times is noticeably larger than the scatter of transit times shown in Figure \ref{fig:orbit_oc}.  Follow-up eclipse observations of HAT-P-13 b would resolve this timing discrepancy.

The depths of our eclipses have shown HAT-P-13 b's atmosphere to be consistent with solar-abundance composition, efficient day-night redistribution, and no thermal inversion layer.  However, we have observed eclipses in only two wavelengths, so further observations in other infrared passbands may indicate more interesting chemistry.

We thank the amateur observers who contributed data to the Exoplanet Transit Database (Table \ref{tab:ttv}).  We thank contributors
to SciPy, Matplotlib, and the Python Programming Language, the free
and open-source community, the NASA Astrophysics Data System, and the
JPL Solar System Dynamics group for software and services.  This work
is based on observations made with the Spitzer Space Telescope, which
is operated by the Jet Propulsion Laboratory, California Institute of
Technology under a contract with NASA.  Support for this work was
provided by NASA through an award issued by JPL/Caltech and through
the NASA Science Mission Directorate's Astrophysics Data Analysis
Program, grant NNX13AF38G, and Planetary Atmospheres Program, grant NNX12AI69G.

\bibliography{hatp13b.bib}

\end{document}